\documentclass[%
 reprint,
 superscriptaddress,
nobibnotes,
 amsmath,amssymb,
 aps,
]{revtex4-2}

\usepackage{graphicx}
\usepackage{dcolumn}
\usepackage{bm}
\usepackage{hyperref}

\usepackage{physics}

\begin{document}

\preprint{APS/123-QED}

\title{Fast Derivative Valuation from Volatility Surfaces using Machine Learning}

\author{Lijie Ding}
\email{dingl1@ornl.gov}
\affiliation{Neutron Scattering Division, Oak Ridge National Laboratory, Oak Ridge, TN 37831, USA}
\author{Egang Lu}
\noaffiliation
\author{Kin Cheung}
\affiliation{Valkin Holdings, LLC, Rowland Heights, CA 91748, USA}
\date{\today}

\begin{abstract}
We introduce a fast and flexible Machine Learning (ML) framework for pricing derivative products whose valuation depends on volatility surfaces. By parameterizing volatility surfaces with the 5-parameter stochastic volatility inspired (SVI) model augmented by a one-factor term structure adjustment, we first generate numerous volatility surfaces over realistic ranges for these parameters. From these synthetic market scenarios, we then compute high-accuracy valuations using conventional methodologies for two representative products: the fair strike of a variance swap and the price and Greeks of an American put. We then train the Gaussian Process Regressor (GPR) to learn the nonlinear mapping from the input risk factors, which are the volatility surface parameters, strike and interest rate, to the valuation outputs.  Once trained, We use the GPR to perform out-of-sample valuations and compare the results against valuations using conventional methodologies. Our ML model achieves very accurate results of $0.5\%$ relative error for the fair strike of variance swap and $1.7\% \sim 3.5\%$ relative error for American put prices and first-order Greeks. More importantly, after training, the model computes valuations almost instantly, yielding a three to four orders of magnitude speedup over Crank-Nicolson finite-difference method for American puts, enabling real-time risk analytics, dynamic hedging and large-scale scenario analysis. Our approach is general and can be extended to other path-dependent derivative products with early-exercise features, paving the way for hybrid quantitative engines for modern financial systems.
\end{abstract}
\maketitle


\section{Introduction}
A financial derivative\cite{hull2016options, shreve2005stochastic} is a contract whose value is derived from the price of an underlying asset, index, or rate. Derivative products such as options\cite{black1973pricing,merton1971theory}, futures, and swaps\cite{flavell2010swaps} have become indispensable tools in modern financial markets, enabling traders and institutions to efficiently transfer and manage risks. Accurate and timely valuation of these instruments underpins effective portfolio risk management\cite{lee2010handbook}, dynamic hedging\cite{taleb1997dynamic} and scenario analysis\cite{hassani2016scenario}. However, the complexity of these products, especially those with path-dependency and early-exercise features, means that closed-form solutions seldom exist and numerical methods\cite{shreve2005stochastic, wilmott1995mathematics} can be prohibitively slow when applied at scale. Moreover, in order to accommodate the single volatility input of the standard Black-Scholes framework, market-observed implied volatilities exhibit pronounced skew and term structure effects \cite{derman2016volatility} forming full curved surfaces in strike and maturity, which add further computational challenge for derivative valuation. As a result, there is a pressing need for fast and flexible pricing engines capable of comprehensive on-the-fly valuation that incorporates such market information.

Recent developments in Machine Learning (ML)\cite{murphy2012machine,carleo2019machine} and Deep Learning\cite{goodfellow2016deep,lecun2015deep} have revolutionized data-driven disciplines across computer science and the physical sciences. In quantitative finance, ML techniques are increasingly being explored to accelerate derivative valuation and risk analysis. Physics-informed neural network (PINN) have been employed to solve option pricing partial differential equation (PDE) directly, embedding the Black–Scholes and similar equations into the network loss function\cite{gatta2023meshless, hainaut2024option, wang2023deep, bai2022application}. Other approaches bypass PDE solvers altogether by learning a direct mapping from model inputs—such as strike, maturity, and underlying volatility—to option prices using feed-forward neural networks\cite{ndikum2020machine, gaspar2020neural, anderson2023accelerated} or Gaussian process regression (GPR) \cite{de2018machine}. However, PINN typically require retraining to accommodate each new set of inputs (e.g., a different strike or volatility level), and existing neural network and GPR schemes generally assume a flat or simplistic volatility structure. As a result, they struggle to generalize across the volatility surfaces observed in real markets.

To overcome these limitations, we have developed a hybrid two-stage valuation framework that applies a ML algorithm on highly accurate derivative valuations that incorporate full volatility surfaces. First, we parameterize the full volatility surfaces using the 5-parameter stochastic volatility inspired (SVI) model\cite{gatheral2011volatility, gatheral2014arbitrage} augmented by a one-factor term structure adjustment, ensuring arbitrage-free behavior across strikes and maturities. From these volatility surface parameters, we then compute high-accuracy valuations via established numerical methods to generate a large dataset of derivatives prices. In the second stage, we train GPR\cite{williams2006gaussian} to learn the nonlinear mapping from the complete set of pricing inputs directly to these valuation outputs. Once trained, the GPR delivers near-instantaneous valuation results. We have applied this methodology on two representative products: variance swap, a forward contract on future realized variance, and American put, an option granting the right to sell the underlying at strike $K$ any time up to expiry. To obtain the valuation results for the dataset, the fair strike $K_{var}$ of variance swap is computed using the standard log-contract replication formula\cite{demeterfi1999more}. As for the American put, the price $V$ plus key Greeks $\Delta$, $\Gamma$ and $\theta$ are obtained via a Crank-Nicolson finite-difference solver. 

The rest of this paper is organized as follows. In Section~\ref{sec:method}, we introduce the SVI model, Crank-Nicolson method and GPR used in this study. We present the results of our study in Section~\ref{sec:results}. Finally, we summarize our paper in Section~\ref{sec:summary}.

\section{Method}
\label{sec:method}

\subsection{Stochastic volatility inspired}
To model the skew and the term structure of the volatility at different strike and maturity, we use the SVI parameterization to construct a smooth volatility surface. Under the SVI model, the total variance $w=\sigma^2_{bs}T$ for maturity $T$ is modeled as a function of the log moneyness $k=\ln{K/F}$, where $\sigma_{bs}$ is the Black-Scholes volatility, $K$ is the strike and $F$ is the forward price. And the total variance is given by \cite{gatheral2014arbitrage}:
\begin{equation}
w(k;\chi_R) = a + b\left[ \rho (k-m) + \sqrt{(k-m)^2 +\sigma^2}\right]
\label{equ: SVI parameterization}
\end{equation}
in which $(a,b,\rho,m,\sigma)$ are the SVI parameters, the total variance need to be positive $w(k)\geq 0$, which requires $a+b\sigma\sqrt{1-\rho^2} \geq 0$. By defining $a' = a + b\sigma\sqrt{1-\rho^2}$, we can decouple the bound to be $a' \geq 0$. And we will denote $\chi_R = (a',b,\rho,m,\sigma)$, whose bounds are given by:

\begin{table}[!ht]
  \centering
  \begin{tabular}{|@{}c| c@{}|}
    \hline
    Parameter bound & Role \\ 
    \hline
    $a'\geq 0$      & vertical translation of the smile \\
    $b\geq0$        & open and close o the smile \\
    $-1<\rho<1$     & counter-clockwise rotation of the smile\\
    $m\in \mathcal{R}$        & horizontal translation of the smile \\
    $\sigma>0$   & at-the-money curvature of the smile \\
    \hline
  \end{tabular}
  \caption{SVI parameter interpretation and admissibility.}
  \label{tab:svi-params}
\end{table}
In reality, the SVI parameters $\chi_R$ can be calibrated based on market data slice-by-slice for each maturity, thus $\chi_R$ is a function of maturity T. In this work, we will only consider a simple one-factor term structure for the variance, such that the shape of the skew are similar across all maturity, and we simply add an additional parameter $\lambda$ and let $\chi'_R = (a',b,\rho,m,\sigma,\lambda)$, assuming:
\begin{equation}
\begin{aligned}
    w(k, T; \chi'_R) &= w(k; \chi_R)f(T; \lambda) \\
    f(T; \lambda) &= T e^{\lambda (1-T)}
\end{aligned}
\end{equation}
such that $w(k, T=1; \chi'_R) = w(k; \chi_R)$ and $w(k, T=0; \chi'_R)=0$, $T\in[0,1]$, $\lambda\in[0,1/T]$. This one-factor term structure is only for illustration purpose and not intended to fit real market data. The arbitrage-free conditions $\pdv*{w(k, T; \chi'_R)}{T} \geq 0$ is then automatically satisfied. Moreover, to find the the local variance $\sigma_{bs}(S, t)$ at time $t$ for stock price $S$ we use the formula \cite{dupire1994pricing, gatheral2011volatility}:
\begin{equation}
    v_L = \frac{\pdv*{w}{T}}{1-\frac{k}{w}\pdv{w}{k}-\frac{1}{4}(\frac{1}{4}+\frac{1}{w}-\frac{k^2}{w^2})(\pdv{w}{k})^2+\frac{1}{2}\pdv[2]{k}{w}}
\end{equation}
and substitute the strike and maturity with stock price and time $\sigma_{loc}^2(S,t) = v_L(k=\log(S/S_0e^{rt}),T=t)$.

\subsection{Crank–Nicolson method}
\label{ssec:Crank–Nicolson}
In order to price the non-dividend American put option and generate training and testing dataset of its theoretical price, we need to solve the Black-Scholes \cite{black1973pricing} partial differential equation with open boundary condition introduced by the early exercise feature. Since there is no analytical solution, we use the Crank–Nicolson\cite{crank1947practical} method, a numerical method widely used in the industry, to numerically solve this. Begin with the Black–Scholes partial differential equation for an option price $V(S,t)$:
\begin{equation}
\pdv{V}{t} + \frac{1}{2}\sigma^2_{loc}(S,t) S^2\pdv[2]{V}{S} + rS\pdv{V}{S} - rV = 0
\end{equation}
where $S$ is the stock price, $t$ is time, $\sigma^2_{loc}(S,t)$ is the local variance \cite{jeanblanc2009mathematical, derman1996local} of the stock at time $t$ and price $S$, $r$ is the interest rate. The boundary condition at the expity $T$ is given by the terminal payoff: 
\begin{equation}
V(S,T)=\Phi(S)=
\begin{cases}
\max(K-S,0),&\text{put}\\
\max(S-K,0),&\text{call}  
\end{cases}
\end{equation}
where $K$ is the strike of the option. In addition, we have appropriate Dirichlet boundary conditions as $S\to0$ and $S\to S_{\max}$ and the free-boundary constraint $V(S,t)\geq\Phi(S)$ due to early exercise. To handle the free-boundary constraint of an American option, we discretize in asset space on a uniform $S$–grid and march backward in time using a Crank–Nicolson \cite{crank1947practical} $\theta$–scheme with Rannacher smoothing. Denote by $V_i^n\approx V(S_i,t_n)$ the option value at node $i$ and time–level $n$.  The semidiscrete $\theta$–scheme reads
\begin{equation}
\frac{V_i^{n+1}-V_i^n}{\Delta t}
+\theta\,\mathcal L\bigl[V^{n+1}\bigr]_i
+(1-\theta)\,\mathcal L\bigl[V^n\bigr]_i
=0    
\end{equation}

where the $\mathcal{L}$ is the spatial operator:
\begin{equation}
\mathcal{L} = \frac{1}{2}\sigma^2_{loc} S_i^2\pdv[2]{S} + rS_i\pdv{S} - r
\end{equation}

By discretization of the spatial operator$\mathcal{L}$, we obtain the Crank-Nicolson update,
\begin{equation}
    \frac{V^{n+1}-V^n}{\Delta t} + \theta\,A\,V^{n+1} + (1-\theta)\,A\,V^n = 0 
\end{equation}
separating the $V^{n+1}$ and $V^{n}$ terms transform our system into a linear system that can be solved using iterations:
\begin{equation}
    \bigl(I + \theta\Delta t A\bigr)\,V^{n+1} = \bigl(I - (1-\theta)\Delta t A\bigr)V^n +\beta^n.
\end{equation}
where $\beta$ is the vector that injects the fixed boundary value. Finally, the free-boundary constraint $V(S,t)\geq\Phi(S)$ is enforced using the Projected Successive over-relaxation algorithm\cite{ikonen2004operator, ikonen2008efficient}.

\subsection{Gaussian Process Regression}

Under the framework of GPR\cite{williams2006gaussian,murphy2012machine}, the goal is to obtain the posterior $p(y_*|\vb{x}_*,\vb{X},\vb{y})$ of the function output $y_* = f(\vb{x}_*)$ with input $\vb{x}_*$, assuming the latent function $f(\vb{x})\sim GP(m(\vb{x},\kappa(\vb{x},\vb{x}'))$ is drawn from a Gaussian process with mean function $m(\vb{x})$ and covariance $\kappa({\vb{x},\vb{x}'})$. Given the training inputs $\vb{X}=(\vb{x}_1, \vb{x}_2, \dots,\vb{x}_n)^T$ and corresponding output $\vb{y}=(y_1,y_2,\dots,y_n)^T$, the joint distribution for the Gaussian process is given by\cite{williams2006gaussian}:
\begin{equation}
    \mqty( \vb{y}\\ y_*)
\sim
\mathcal{N} \Biggl( \vb{0},
\mqty[
\kappa(\vb{X},\vb{X}) + \sigma_g^2 \vb{I} & \kappa(\vb{X},\vb{x}*)\\
\kappa(\vb{x}_*,\vb{X}) & \kappa(\vb{x}_*,\vb{x}_*)
] \Biggr)
\end{equation}
where $\vb{I}$ is the unit matrix, and we use the Radial basis function (Gaussian) kernel $\kappa(\vb{x},\vb{x}')=\exp(-|\vb{x}-\vb{x}'|^2/2l_g)$. The length scale $l_g$ and noise level $\sigma_g$ are the hyperparameters to be optimized using the training data by maximizing the log marginal likelihood\cite{williams2006gaussian}:
\begin{equation}
\log p(\vb{y}|\vb{X}) = -\tfrac12 \vb{y}^T \mathcal{K}^{-1}\vb{y} -\tfrac12\log \det \mathcal{K} -\frac{n}{2}\log(2\pi)
\end{equation}
where $\mathcal{K} = \kappa(\vb{X},\vb{X})$ and $n$ is the number of training data. Conditioning on the observations of training data, the posterior predictive distribution is Gaussian:
\begin{equation}
\begin{aligned}
    &p(y_*|\vb{x}_*,\vb{X},\vb{y}) = \mathcal{N}(\mu_*,\sigma_*)\\
    &\mu_* = \kappa(\vb{x}_*,\vb{X})\bigl[\mathcal{K} + \sigma_g^2 \vb{I}\bigr]^{-1}\vb{y} \\
    &\sigma^2_* = \kappa(\vb{x}_*,\vb{x}_*) - \kappa(\vb{x}_*,\vb{X})\bigl[\mathcal{K} + \sigma_g^2 \vb{I}\bigr]^{-1}\kappa(\vb{X},\vb{x}_*) \\
\end{aligned}
\end{equation}

For our calculation of variance swap and American put, we just need to substitute appropriate $\vb{x}$ and $y$ into the GPR. For instance, the valuation of variance swap is all about the calculation of the fair strike $y_*=K_{var}$, whose risk factor includes the SVI skew parameters and the interest rate $\vb{x} = (a',b, \rho,m,\sigma,r)$. As for the American put, the input involves the whole SVI surface parameter, the strike, and the interest rate $\vb{x}=(a',b, \rho,m,\sigma,\lambda,K, r)$, and we will attempt to predict not only the option price $V$, but also the Greeks Delta, Gamma and Theta, such that $y_*\in (V,\Delta,\Gamma,\Theta)$. Finally, we utilize
the scikit-learn \cite{pedregosa2011scikit, buitinck2013api} Gaussian Process library in practice, due to its convenience and efficiency.

\section{Results}
\label{sec:results}
We start with illustrating the volatility surface modeled by the SVI model, demonstrating the effect of each SVI parameter and term structure parameters on the volatility skew and surface. Then we discuss the pricing of variance swap as a function of the volatility skew, generate training and testing data for the ML model, and demonstrate the effectiveness of the pricing of variance swap using GPR. Finally, we apply the similar approach for the American put option, whose valuation depends on the entire volatility surface, and show the applicability of our framework for the pricing of American put.  In our study, we consider the natural unit of maximum maturity $T_{max}=1$ in the volatility surface, and only consider the price of variance swap and American put with maturity $T=1$ and stock spot price $S_0=S(t=0)=1$.


\subsection{Volatility Surface}

\begin{figure}[!th]
    \centering
    \includegraphics[width=\linewidth]{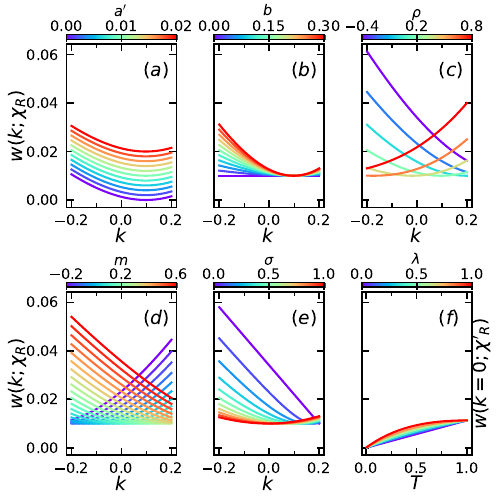}
    \caption{Variation of variance skew and term structure for 6 variance parameters $\chi'_R$, the based line parameter is $\chi'_R=(a',b,\rho,m,\sigma, \lambda)=(0.01, 0.15, 0.2, 0.2, 0.5, 0)$, each plot only vary one parameter at a time. (a-e) Variance skew of total variance $w$ versus log moneyness $k$ at $T=1$ for various $a'$, $b$, $\rho$, $m$ and $\sigma$, respectively. (f) Term structure of the at-the-money total variance versus maturity $T$ for different term structure parameter $\lambda$.}
    \label{fig:svi_curve}
\end{figure}

\begin{figure}[!ht]
    \centering
    \includegraphics[width=\linewidth]{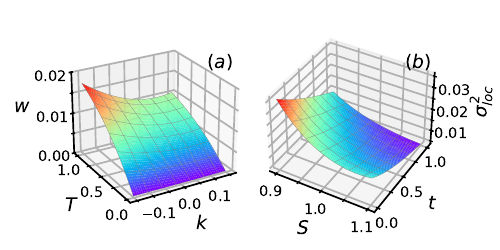}
    \caption{Example of the variance surface and the derived local variance at $\chi'_R=(0.01, 0.15, 0.2, 0.2, 0.5, 0.5)$ where $\chi'_R=(a',b,\rho,m,\sigma,\lambda)$. (a) Total variance surface $w(T,K)$ as function of maturity $T$ and strike $K$. (b) Local variance surface $\sigma^2_{loc}(S,t)$ as the function of local stock price $S$ and time $t$. The gray wire frame is only for visual guide.}
    \label{fig:svi_surface}
\end{figure}

To illustrate the effect of each SVI parameters on the volatility surface, we plot the total variance $w(k;\chi_R)$ as a function of the log moneyness $k$ for various $\chi_R$. As shown in Fig.~\ref{fig:svi_curve} (a-e), each parameter $(a',b,\rho,m,\sigma)\in\chi_R$ has significant effect on the shape of the skew of the total variance $w(k;\chi_R)$. Consistent with the description of the SVI model\cite{gatheral2014arbitrage}, $a'$ control the vertical shift of the skew, $b'$ tunes the open and close of the smile, with $b=0$ correspond to a flat volatility skew. $\rho$ rotate the entire skew, which rotate counter-clockwise as $\rho$ increases. $m$ translates the skew horizontally towards the right direction of $k$ as it increases, moving the left side of the volatility smile into the $k$ window, Finally, the $\sigma$ control the curvature of the skew. In addition, the term structure of the at-the-money total variance $w(k=0)$ is shown in Fig.~\ref{fig:svi_curve}(f), which goes to $w(T=0)=0$ as the maturity become zero, while the term structure parameter $\lambda$ control the rate of decay from $w(T=1)$ to $w(T=0)$. 

Moreover, Fig.~\ref{fig:svi_surface}(a) shows an example of the entire total variance surface $w(k,T)$ while Fig.~\ref{fig:svi_surface}(b) shows the local variance $\sigma^2_{loc}(S,t)$ derived from the total variance surface.

\subsection{Variance Swap}

The valuation of the variance swap breaks down to the calculation of the fair strike, which is given by: \cite{demeterfi1999more}
\begin{equation}
\begin{aligned}
    & K_{var} = \frac{2}{T} \biggl[ rT + 1 - e^{rT}  \\
              & +e^{rT} \biggl( \int_0^{S_0} \frac{P(K)}{K^2}  \dd{K} + \int_{S_0}^{\infty} \frac{C(K)}{K^2}  \dd{K} \biggr) \biggr]
\end{aligned}
\label{eq:variance_swap_Kvar}
\end{equation}

where $r$ is the interest rate, $T=1$ is the maturity, $S_0=1$ is the spot price and $C(K)$ and $P(K)$ denote the price of the European call and put option of strike $K$, given by\cite{hull2016options, black1973pricing}:
\begin{equation}
\begin{aligned}
C(K) &= S_0 N(d_1) - K e^{-rT} N(d_2) \\
P(K) &= K e^{-rT} N(-d_2) - S_0 N(-d_1) \\
d_1 &= \left[\log(S_0/K) + \left(r + \sigma^2_{bs}/2 \right)T \right]/\sigma_{bs} \sqrt{T} \\
d_2 &= d_1 - \sigma_{bs} \sqrt{T}
\end{aligned} 
\label{eq:BS_price}
\end{equation}
where $N(\cdot)$ denotes the standard normal cumulative distribution function. The Black-Scholes variance $\sigma_{bs}(K)$ for different strike is given by the SVI total variance $\sigma_{bs}(K)=w(\log{K/S_0e^{rT}} ,\chi_R)$. Substituting all necessary terms into the Eq.~\eqref{eq:variance_swap_Kvar} provides the fair strike of the variance swap $K_{var}$ for the given variance skew defined by the SVI parameter $\chi_r$ as well as the interest rate $r$.

\begin{figure}[!th]
    \centering
    \includegraphics[width=\linewidth]{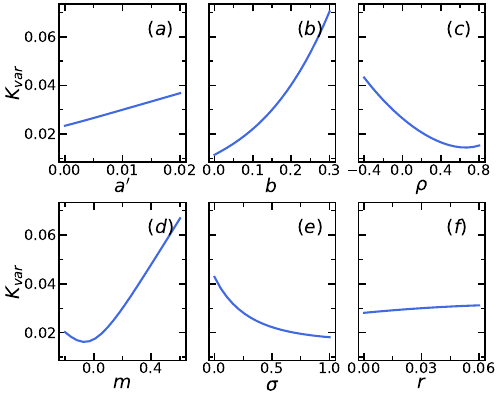}
    \caption{Variation of the fair strike of variance swap for various risk factors including variance skew parameters and interest rate. The based line parameters are $r=0.03$ and $\chi_R=(a',b,\rho,m,\sigma)=(0.01, 0.15, -0.1, 0.2, 0.2)$, corresponding to $\sigma_{bs}=0.16$ at-the-money Black-Scholes volatility. Only one parameter is varied at every plot each time. (a-e) Fair strike $K_{var}$ versus various SVI parameters in $\chi_{R}$. (f) Fair strike versus interest rate.}
    \label{fig:Kvar_vs_params}
\end{figure}

To inspect the effect of each risk factor for the fair strike of the variance swap, including the SVI parameter $\chi_R$ and interest rate $r$, we investigate the variation of fair strike $K_{var}$ as each risk factor is tuned independently. As shown in Fig.~\ref{fig:Kvar_vs_params}, the fair strike $K_{var}$ vary smoothly in along all directions of these risk factor, indicate the feasibility of mapping the risk factors $(a',b,\rho,m,\sigma,r)$ to fair strike $K_{var}$ using GPR. The range of $K_{var}$ is just for illustration of the sensitivity analysis centering at the specific $r=0.03$ and $\chi_R=(a',b,\rho,m,\sigma)=(0.01, 0.15, -0.1, 0.2, 0.2)$, which correspond to the at-the-money Black-Scholes volatility of $\sigma_{bs}=0.16$. This reference point is chosen purely for convenience; any other center would serve just as well and would not affect the qualitative conclusion.

\begin{table}[!b]
    \centering
    \begin{tabular}{ | l | l | l |}
    \hline
         & train & test \\ \hline
         $a'$ & $U(0, 0.02)$ & $U(0.005,0.015)$ \\ 
         $b$ & $U(0, 0.3)$ & $U(0.05,0,25)$\\
        $\rho$ & $U(-0.4, 0.8)$ & $U(-0.3,0.7)$ \\
        $m$ & $U(-0.2, 0.6)$ & $U(-0.1, 0.5)$ \\
        $\sigma$ & $U(0,1)$ & $U(0.1,0.9)$ \\
        $r$ & $U(0,0.06)$ & $U(0.01,0.05)$  \\
    \hline
    \end{tabular}
    \caption{Parameter range for the SVI parameter $\chi_R$ and interest rate $r$ for the generating the dataset of variance swap. $U(a,b)$ is the uniform distribution within the interval $[a,b]$.}
    \label{tab:vs_parameter_range}
\end{table}

To put the pricing of the variance swap using GPR into practice, we prepared the training data and testing data by randomly sampling the risk factors $(a', b,\rho,m,\sigma,r)$ from the uniform distribution shown in Tab.~\ref{tab:vs_parameter_range}, and then calculate the corresponding fair strike $K_{var}$ numerically based on Eq.~\eqref{eq:variance_swap_Kvar}. There are $2,000$ training data and $2,000$ testing data we generate in total. We then train the GPR using the testing data by maximizing the log marginal likelihood for the training date respect to the length scale hyperparameter $l_g$. We note that while in general there is also have a noise level hyperparameter $\sigma_g$, but for the case of variance swap, we found $\sigma_g < 10^8$ is very small, thus we simply set it to zero, and only keep the length scale $l_g$. 

\begin{figure}[!h]
    \centering
    \includegraphics[width=\linewidth]{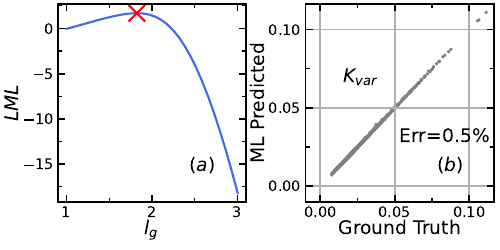}
    \caption{Predict the fair strike using Gaussian Process Regressor (GPR). (a) Log marginal likelihood (LML) of the test data, normalized by the number of data, versus correlation length $l_g$. Optimized $l_g$ is highlighted with red cross mark. (b) Comparison between the ground truth fair strike $K_{var}$ in the test data and the Machine Learning predicted value by passing the pricing parameters $(\chi_R, r)$ into the trained GPR. Relative error are annotated in the plot.}
    \label{fig:variance_swap_GPR_fitting}
\end{figure}

Fig.~\ref{fig:variance_swap_GPR_fitting}(a) shows the normalized log marginal likelihood $LML=p(\vb{y}|\vb{X})/n$, where for the case of variance swap $y=K_{var}$, $\vb{x} = (a',b,\rho,m,\sigma,r)$ and $n=2,000$. We then apply the trained GPR to predict the fair strike $K_{var}$ for the data from the testing data. Fig.~\ref{fig:variance_swap_GPR_fitting}(b) shows the comparison between the ground truth $K_{var}$ calculated from the Eq.~\ref{eq:variance_swap_Kvar} for $\vb{x_*}$ from the testing data and the ML predicted value from the trained GPR. Defining the relative error $Err(\zeta) = \left<|\zeta'-\zeta|\right>/\left<\zeta\right>$ for valuation parameter $\zeta$ with predicted value $\zeta'$. Considering the sparsity of $2,000$ points in the 8-dimensional parameter space, we only have very small number of testing data. Nevertheless, the ground truth and ML predicted $K_{var}$ agree very well, with relative error $Err(K_{var})=0.5\%$, demonstrating the effectiveness of our GPR approach for the variance swap.

\subsection{American Put}

We then move on to apply the same approach to the pricing of American put option. We first discuss the solution of American put, demonstrate the significance of it's pricing, then investigate the effect of each risk factors on the price and Greeks of the American put, Finally, we generate the training set and testing set consisting of random combinations of the risk factors, including all of 6 SVI variance surface parameters $\chi'_R=(a',b,\rho,m,\sigma,\lambda)$, the strike $K$, and interest rate $r$, and corresponding pricing results, then illustrate the effectiveness of the ML pricing approach.

\begin{figure}[!ht]
    \centering
    \includegraphics[width=\linewidth]{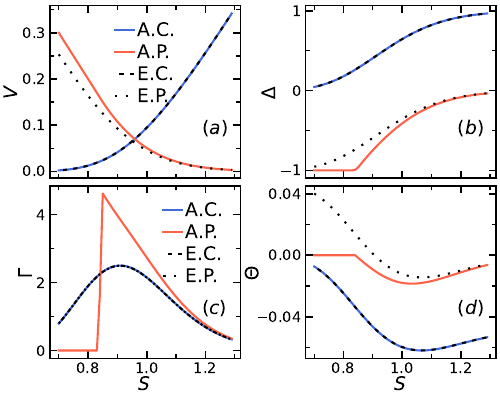}
    \caption{Solution of the option price and Greeks at $t=0$ as function of stock price $S$ for all four types of options: American call (A.C.), American put (A.P.), European call (E.C.) and European put (E.P.) with flat volatility $\sigma^2_{loc} = 0.03$, interest rate $r=0.05$, and strike $K=1$. (a) Option price $V$. (b) Option Delta $\Delta=\pdv{V}{S}$. (c) Option Gamma $\Gamma=\pdv[2]{V}{S}$. (d) Option Theta $\Theta=\pdv{V}{t}$.}
    \label{fig:american_price_solution}
\end{figure}

While the price for the European options can be analytically calculated using Eq.~\ref{eq:BS_price}, there is no analytical solution for the American put due to early exercise and open-boundary condition. The solution is obtained numerically using the Crank-Nicolson method introduced in Sec.~\ref{ssec:Crank–Nicolson}. Fig.~\ref{fig:american_price_solution} shows the solution for price $V(S)$ and Greeks, including Delta $\Delta(S) = \pdv{V}{S}$, Gamma $\Gamma(S) = \pdv[2]{V}{S}$ and Theta $\Theta(S)=\pdv{V}{t}$ as a function of the spot price $S$ at time $t=0$ for a flat volatility surface. As shown in Fig.~\ref{fig:american_price_solution}(a) the American call has the same solution as the European call, making it's pricing relatively trivial. Meanwhile, the price of American put is higher than the European call especially when it is in-the-money (spot lower than strike $S<K$). This is due to the early exercise feature of the American put. Such consequence is also reflected in the Greeks. Notably, the early exercise results in a discontinuity for the second derivative Gamma, as shown in Fig.~\ref{fig:american_price_solution}(c). Due to this discontinuity in the spot dimension, we expect a similar discontinuity appears in the strike $K$ dimension when calculating the Gamma for spot price $S_0=1$. Such jump will affect the effectiveness of the pricing using GPR, as GPR is based on the continuity of the mapping function.

\begin{figure}[!ht]
    \centering
    \includegraphics[width=\linewidth]{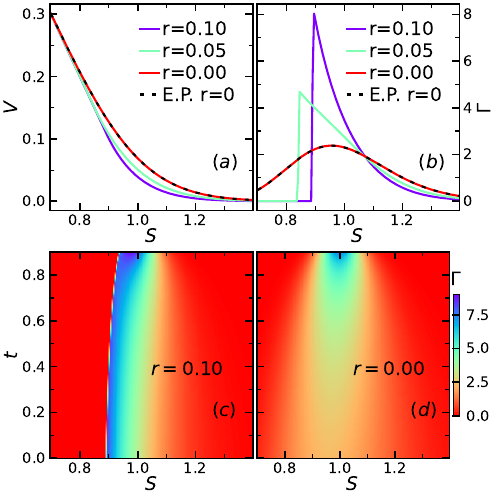}
    \caption{Entire solution of American put price $V$ and Gamma $\Gamma$ for various interest rate $r$ with flat volatility $\sigma^2_{loc} = 0.03$ and strike $K=1$. There is a cutoff time on $\Gamma$ since it diverges at $t=1$. (a) American put option price $V(S)$ at $t=0$ for various interest rate, dashed line is European put with zero interest $r=0$. (b) American put option Gamma $\Gamma(S)$ at $t=0$ for various interest rate, dashed line is European put with zero interest $r=0$. (c) Heat map of the option Gamma as function of time and stock price $\Gamma=\Gamma(S,t)$ for interest rate $r=0.1$. (d) Similar to (c) but with zero interest rate $r=0$.}
    \label{fig:american_price_solution_per_r}
\end{figure}

An intuitive explanation of the discontinuity is that, even for a non-dividend stock, when the stock price is low enough, the room for further price drop is limited, cap the time value of the option. On the other hand, the holder of American put can simply exercise the option, take the payoff and earn the interest. When the interest from the payoff is greater than the time value of the put option, it is optimal to exercise. Fig.~\ref{fig:american_price_solution_per_r}(a) and (b) show the price $V(S)$ and Gamma $\Gamma(S)$ at $t=0$, for different interest rate. The $r=0$ curves reduce to the solution of European put. Fig.~\ref{fig:american_price_solution_per_r}(c) and (d) show the solution of $\Gamma(S,t)$ for interest rate $r=0.1$ and $r=0$, respectively. The sharp color change in Fig.~\ref{fig:american_price_solution_per_r}(c) indicate the the early exercise, which is absent in the zero interest $r=0$ case as shown in Fig.~\ref{fig:american_price_solution_per_r}(d).

\begin{figure}[!ht]
    \centering
    \includegraphics[width=\linewidth]{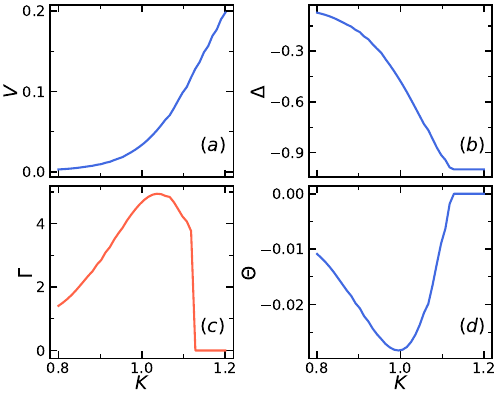}
    \caption{Variation of the option price and Greeks for the American put versus strike $K$ for SVI parameters $\chi'_R = (a',b,\rho,m,\sigma,\lambda)=(0.01, 0.15, 0.2, 0.2, 0.5, 0.5)$ and interest rate $r=0.03$, (a) Option price $V$ versus each pricing parameters. (b) Option Delta $\Delta=\pdv{V}{S}$. (c) Option Gamma $\Gamma=\pdv[2]{V}{S}$. (d) Option Theta $\Theta=\pdv{V}{t}$.}
    \label{fig:american_params_sensitivity_price}
\end{figure}

We then investigate the sensitivity of the price and Greeks of the American put on the different risk factors especially the occurrence of discontinuity. Fig.~\ref{fig:american_params_sensitivity_price} shows the variance of $(V,\Delta,\Gamma,\Theta)$ respect to strike $K$. As expected, Fig.~\ref{fig:american_params_sensitivity_price}(c) shows discontinuity of Gamma $\Gamma$ versus strike $K$, imply difficulty for GPR calculation of Gamma $\Gamma$. More detail about the continuity of the price and Greeks versus other risk factors in $(a',b,\rho,m,\sigma,\lambda,t)$ can be found in the Appx.~\ref{sec:sensivity analysis}.

\begin{table}[!h]
    \centering
    \begin{tabular}{ | l | l | l |}
    \hline
         & train & test \\ \hline
         $K$ & $U(0.85, 1.15)$ & $U(0.9, 1.1)$ \\ 
         $\lambda$ & $U(0, 1)$ & $U(0.1,0,9)$\\
    \hline
    \end{tabular}
    \caption{Additional parameter range for American put, such as strike $K$ and term structure parameter $\lambda$.}
    \label{tab:ap_parameter_range}
\end{table}

\begin{figure}[!ht]
    \centering
    \includegraphics[width=\linewidth]{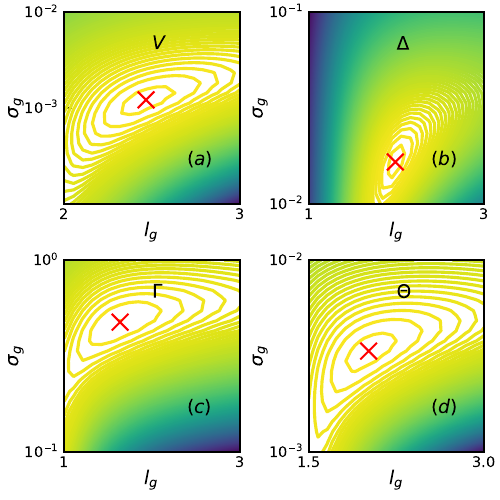}
    \caption{Log marginal likelihood of the training data versus two hyperparameters, length scale $l_g$ and noise level $\sigma_g$, for option price and Greeks. (a) Option price $V$. (b) Option Delta $\Delta=\pdv{V}{S}$. (c) Option Gamma $\Gamma=\pdv[2]{V}{S}$. (d)  Option Theta $\Theta=\pdv{V}{t}$. The optimized $l_g$ and $\sigma_g$ are highlighted using red cross mark. The absolute value the LML is not important.}
    \label{fig:american_put_LML}
\end{figure}

To generate the training set and testing set, we use parameters range same as in Tab.~\ref{tab:vs_parameter_range} for $(a',b,\rho,m,\sigma,r)$, and the range for strike $K$ and term structure parameter $\lambda$ are given by Tab.~\ref{tab:ap_parameter_range}. The choice of parameter range result in the range of at-the-money Black-Scholes volatility $\sigma_{bs}(k=0)\sim (0.003, 0.5)$. We prepare $5,000$ data for the training set and $2,000$ for the testing set based on the give range of parameters. Similar to the procedure for variance swap, we firstly find the optimized hyperparameters $(l_g, \sigma_g)$ by maximizing the log marginal likelihood (LML) of the testing set. Fig.~\ref{fig:american_put_LML} shows the the contour of the LML for each target: price $V$, Delta $\Delta$, Gamma $\Gamma$ and Theta $\Theta$. 

\begin{figure}[!ht]
    \centering
    \includegraphics[width=\linewidth]{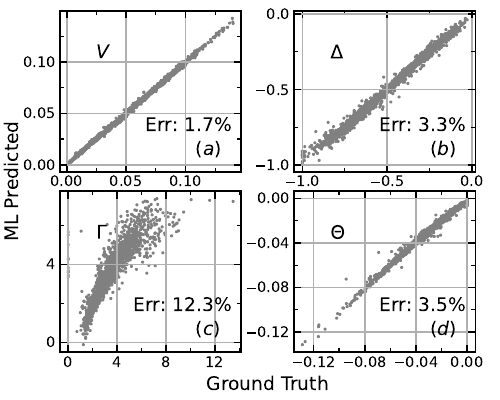}
    \caption{Comparison between Machine Learning predicted option price and Greeks and the ground truth for American put option. Relative error are annotated in each plots. (a) Option price $V$. (b) Option Delta $\Delta=\pdv{V}{S}$. (c) Option Gamma $\Gamma=\pdv[2]{V}{S}$. (d)  Option Theta $\Theta=\pdv{V}{t}$.}
    \label{fig:american_put_GPR_fitting}
\end{figure}

After the GPR is trained for the American put, we apply the GPR to find the price and Greeks of the testing data. Fig.~\ref{fig:american_put_GPR_fitting} shows the comparison between the ground truth of the testing data, solved by the Crank-Nicolson method, and the corresponding ML prediction. The ML prediction for price $V$, Delta $\Delta$ and Theta $\Theta$ are relatively good, while Gamma $\Gamma$ deviate a lot when the ground truth $\Gamma$ is large, which corresponds to the jump of $\Gamma$ from large value to zero. This discrepancy reflect the inherent limitation of the GPR method, as it implicitly assumes that the output is continuous respect to the variation of input variables. Nevertheless, the first-order Greeks are relative accurate and the the price prediction achieved high accuracy with $1.7\%$ relative error, despite of the small number of data in our training set, considering the 8-dimensional input space we have for the American put.

Finally, the ML approach offers tremendous efficiency improvement in terms of the computational time required for finding the price and Greeks. Tab.~\ref{tab:run_time} compare the CPU time for pricing 2,000 American put using Crank-Nicolson with different grid size and the ML prediction. While the CPU time for the Crank-Niclson method increases as the grid size become smaller, pricing massive amount of trades still require significant amount of CPU time. Meanwhile, the GPR, once trained, can find the price and Greeks of the American put almost instantly.

\begin{table}[!h]
    \centering
    \begin{tabular}{|c |c |c |c |c|}
    \hline
    & Machine Learning & \multicolumn{3}{c|}{Crank-Nicolson}  \\
    \hline
     Grid & N/A  &  $200^2$ & $500^2$ &$1000^2$  \\ \hline
      Time & 2.09s &  2,386.08s & 17,966.34s & 61,198.82s    \\ \hline 
      Speedup & $1\times$  & $1141\times$ & $8596\times$ & $29281\times$ \\ \hline
    \end{tabular}
    \caption{Running time for finding the price and Greeks of all 2,000 American put. Tested on Apple M2 CPU. CPU time for Crank-Nicolson are estimated by running the solver for 5 minutes, count the number of solution then scale to the full 2,000 data.}
    \label{tab:run_time}
\end{table}

\section{Summary}
\label{sec:summary}
In this work, we introduce a ML framework that takes all the relevant risk factors as input, including the parameters modeling the shape of the volatility surface, and generates the price and related Greeks as output directly almost instantly. To illustrate this methodology, we have used idealized volatility surfaces where the volatility surface for any given maturity is described by the  5-parameter SVI parameterization, and the term structure is specified by a single parameter. Within this idealized framework, we then apply this methodology to evaluate two kinds of derivatives products, namely the fair strike $K_{var}$ of a variance swap, and the price $V$ and Greeks $(\Delta,\Gamma,\Theta)$ of an American put. For each of these products, we have prepared a training set and a testing set using valuations obtained by highly accurate numerical models commonly used by derivatives practitioners. The training data are then used to train a GPR to learn the mapping between the input risk factors and the output valuation variables directly, and the performance of the GPR is validated using the testing data where the high accuracy numerical model valuations are used as the ground truth. For the variance swap, a very high precision prediction with an overall $0.5\%$ relative error is achieved. As for the American put, the price $V$ and first order Greeks $\Delta$ and $\Theta$ all have accurate predictions with relative error at $1.7\%$, $3.3\%$ and $3.5\%$ , respectively. However, partly due to the discontinuity of the Gamma $\Gamma$ profile in the strike dimension, the GPR's performance of this higher order derivative valuation  is notably less accurate. Nonetheless, the key message from this study is that by training ML to directly map the relationships between pricing inputs and valuation outputs, this methodology has reduced the computation time by 3 to 4 orders of magnitude for the American put, offering significant improvement and potential in performing large scale real-time valuations of derivative products with early exercise features.

Although we have only demonstrated the application of our methodology on two relatively simple products in this work, the principal contributions lie in the ability to incorporate full volatility surface information as well as the near instantaneous valuation. In addition, it is worth noting that our framework is extremely flexible as it is completely agnostic on the type of valuation engine being used to generate the training data set. For example, one can easily replace the Crank-Nicolson valuation model by one's own proprietary pricing engine to maintain consistency with existing valuation systems. Moreover, our approach should be sufficiently general that it can be applied to other more complicated path dependent exotic derivatives, for example, Asian option\cite{rogers1995value}, Bermudan option\cite{schweizer2002bermudan, longstaff2001valuing}, Binary and Barrier options, as well autocallables \cite{guillaume2015autocallable} with knock-in features, etc. The significant improvement in pricing efficiency will hopefully allow market participants to perform real-time risk management and scenario analysis, as well as embed these pricing routines into larger calibration and optimization workflows. 

Looking forward, there are three key directions we plan to pursue to strengthen and extend this framework. First, we can explore alternative regression techniques to overcome the challenges posed by valuation discontinuities. Second, we aim to enrich our term-structure modeling by allowing all 5 SVI parameters to vary dynamically over time, capturing the wider range of the evolving smile and skew of real-market surfaces. Most importantly, we will need to transition from synthetically generated data to real-world market data for training and testing our methodology.

\section{Data Availability}
The code for the numerical calculation and data analysis are available at the GitHub repository \href{https://github.com/ljding94/GPR_pricing}{GPR pricing}

\section{Author Contributions}
LD, EL and KC conceived the work; LD derived the theoretical framework, developed the code, generated and analyzed the data; and LD, EL, and KC wrote and edited the manuscript.

\section{Acknowledgement}
This research was performed at the Spallation Neutron Source, which is DOE Office of Science User Facilities operated by Oak Ridge National Laboratory. This research was sponsored by the Laboratory Directed Research and Development Program of Oak Ridge National Laboratory, managed by UT-Battelle, LLC, for the U.S. Department of Energy. This research used resources of the Compute and Data Environment for Science (CADES) at the Oak Ridge National Laboratory, which is supported by the Office of Science of the U.S. Department of Energy under Contract No. DE-AC05-00OR22725. 


\bibliography{reference}

\onecolumngrid
\clearpage
\appendix
\setcounter{figure}{0}
\renewcommand{\thefigure}{A\arabic{figure}}
\section{Sensitivity analysis for American put}
\label{sec:sensivity analysis}
\begin{figure}[!h]
    \centering
    \includegraphics[width=\linewidth]{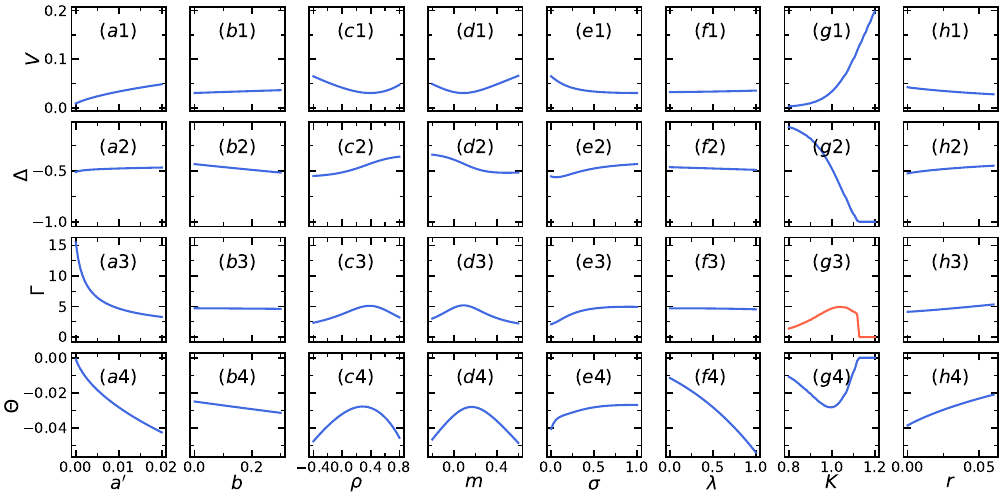}
    \caption{Variation of the option price and Greeks for the American put versus SVI parameters $\chi'_R$, interest rate $r$ and strike $K$. The based line parameters are $\chi'_R=(a',b,\rho,m,\sigma,\lambda)=(0.01, 0.15, 0.2, 0.2, 0.5, 0.5)$, $r=0.03$ and $K=1$. Only one parameter is varied at every plot each time. (a1-h1) Option price $V$ versus each pricing parameters. (a2-h2) Option Delta $\Delta=\pdv{V}{S}$. (a3-h3) Option Gamma $\Gamma=\pdv[2]{V}{S}$. (a4-h4) Option Theta $\Theta=\pdv{V}{t}$.}
    \label{fig:american_params_sensitivity_price_all}
\end{figure}

\end{document}